\pdfoutput=1

\documentclass[11pt,table]{article}

\usepackage[]{ACL2023}

\usepackage{algorithm}
\usepackage{algorithmic}
\algsetup{indent=0.5em} 
\usepackage{times}
\usepackage{latexsym}

\usepackage[T1]{fontenc}

\usepackage[utf8]{inputenc}
\usepackage{listings}
\usepackage{xcolor} 
\usepackage{microtype}
\usepackage{threeparttable}
\usepackage{tablefootnote}
\usepackage{booktabs}
\usepackage{microtype}
\usepackage{amsmath}
\usepackage{amsfonts} 
\usepackage{multirow}
\usepackage{xcolor}
\usepackage{tabularx}
\usepackage{makecell}
\usepackage{inconsolata}
\usepackage{graphicx}
\usepackage{tikz}
\usepackage{tikz-qtree}
\usetikzlibrary{trees}
\usepackage{forest}
\usetikzlibrary{fit}

\usepackage{xcolor}
\usepackage{tcolorbox}
\usepackage{multirow}

\definecolor{mygray}{RGB}{102,102,102}
\definecolor{myred}{RGB}{178,34,34}
\usepackage{inconsolata}

\usepackage{graphicx}
\usepackage{amsmath}
\usepackage{amsfonts}
\usepackage{amssymb}

\usepackage{booktabs}  
\usepackage{pifont}    
\usepackage{graphicx}  
\usepackage{amssymb}   
\usepackage{hyperref}  

\definecolor{color1}{RGB}{232,232,232}
\definecolor{color2}{RGB}{255,240,245}
\definecolor{color3}{RGB}{224,255,255}
\definecolor{color4}{RGB}{255, 235, 178}

\definecolor{mygray}{RGB}{102,102,102}
\definecolor{myred}{RGB}{178,34,34}

\title{InteractiveSurvey: An LLM-based Personalized and Interactive Survey Paper Generation System}

\author{Zhiyuan Wen\textsuperscript{1}, Jiannong Cao\textsuperscript{1}, Zian Wang*\textsuperscript{1}, Beichen Guo*\textsuperscript{1}, Ruosong Yang\textsuperscript{1}, Shuaiqi Liu\textsuperscript{2}\\
  \textsuperscript{1}The Hong Kong Polytechnic University, Kowloon, Hong Kong, China\\
  \textsuperscript{2}Huawei Technologies Co., Ltd, Shenzhen, China\\
  \texttt{\{zyuanwen, jiannong.cao\}@polyu.edu.hk} \\
  \texttt{\{atopos.wang,beichen.guo\}@connect.polyu.hk} \\
  \texttt{rsong.yang@polyu.edu.hk} \\
  \texttt{liushuaiqi@huawei.com}\\
}

\begin{document}
\maketitle

\begin{abstract}

The exponential growth of academic literature creates urgent demands for comprehensive survey papers, yet manual writing remains time-consuming and labor-intensive. Recent advances in large language models (LLMs) and retrieval-augmented generation (RAG) facilitate studies in synthesizing survey papers from multiple references, but most existing works restrict users to title-only inputs and fixed outputs, neglecting the personalized process of survey paper writing. In this paper, we introduce InteractiveSurvey - an LLM-based personalized and interactive survey paper generation system. InteractiveSurvey can generate structured, multi-modal survey papers with reference categorizations from multiple reference papers through both online retrieval and user uploads. More importantly, users can customize and refine intermediate components continuously during generation, including reference categorization, outline, and survey content through an intuitive interface. Evaluations of content quality, time efficiency, and user studies show that InteractiveSurvey is an easy-to-use survey generation system that outperforms most LLMs and existing methods in output content quality while remaining highly time-efficient\footnote{Our demo video is at: \href{https://technicolorguo.github.io/InteractiveSurvey/}{here}}.

\end{abstract}

\section{Introduction}

\begin{figure}[t]
  \centering
  \includegraphics[scale=0.35]{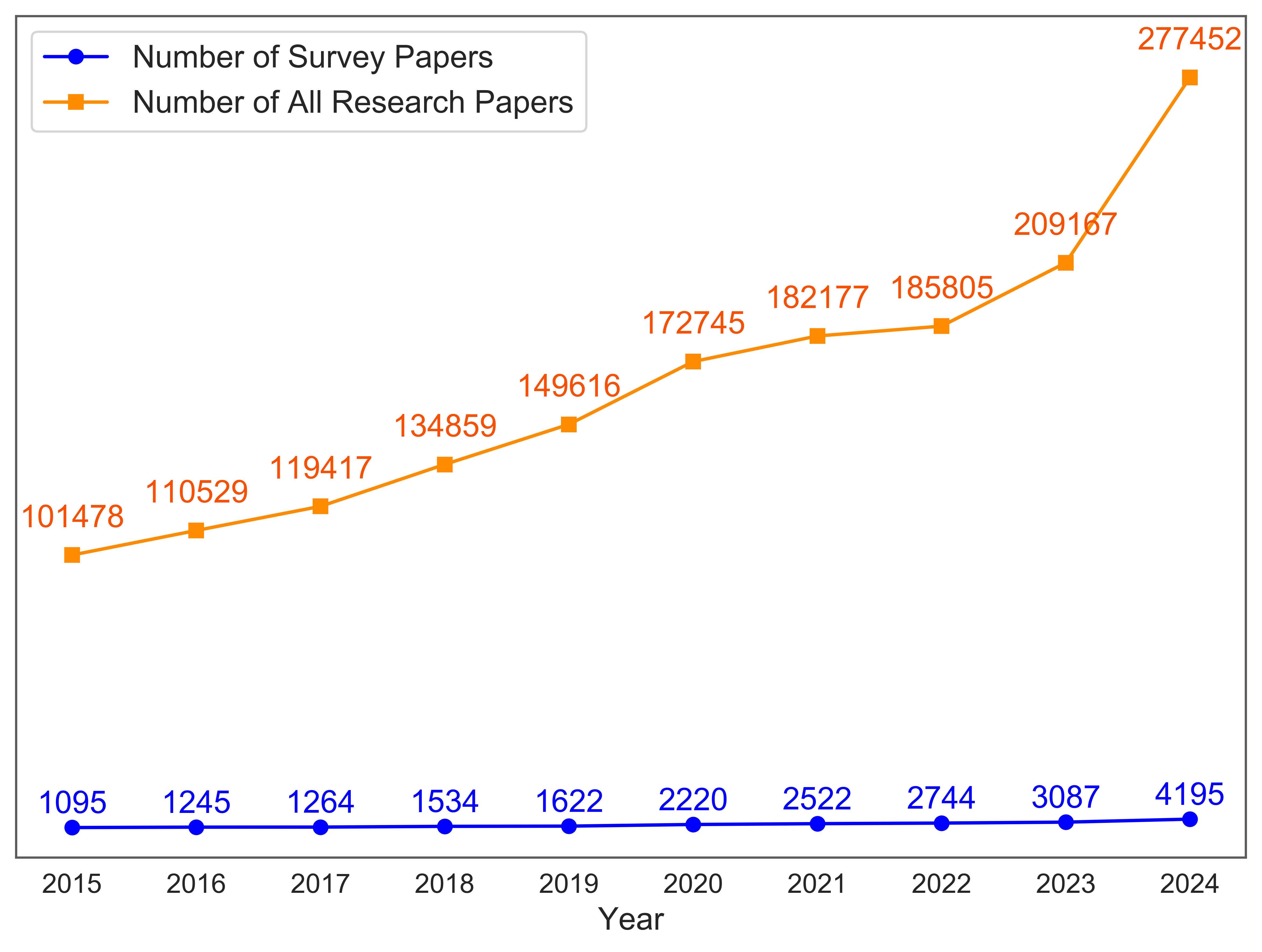}
  \caption{Comparison of the number of all research papers and survey papers released on arXiv.org over the past 10 years (2015–2024).}
  \label{fig:intro}
\end{figure}

Survey papers are essential for synthesizing the current state and trends of a research area. While research papers have grown rapidly over the past decade, survey papers remain relatively scarce (Figure \ref{fig:intro}). This gap makes it challenging to keep up with developments in a field \cite{wang_autosurvey_2024}, especially for new researchers. Consequently, there is an urgent demand to generate high-quality survey papers efficiently.

Generating a comprehensive survey typically involves summarizing dozens to hundreds of references, with each averaging around 10K tokens, far exceeding the input capacity of most mainstream LLMs like GPT-4o \cite{hurst2024gpt}.
Recent advances in retrieval-augmented generation (RAG, \citet{gao2023retrieval}) facilitate synthesizing survey papers from multiple references \cite{wang_autosurvey_2024,liang2025surveyx,torres_promptheus_2024,agarwal2024litllmtoolkitscientificliterature}. However, most existing approaches/systems restrict users to title-only inputs and fixed outputs, failing to involve them in the intermediate stages of survey writing, such as selecting and categorizing references or modifying survey paper outlines. Consequently, if users are dissatisfied with certain sub-parts, they have to either regenerate the entire survey paper or are even unable to adjust the content. This significantly prevents the usability and efficiency of survey paper generation.

In this paper, we introduce InteractiveSurvey, an LLM-based interactive web system that can efficiently generate personalized and comprehensive survey papers for researchers. InteractiveSurvey has the following functions and features: \textbf{(1)  Automatic Reference Searching}: our system automatically searches and downloads reference papers from arXiv that are relevant to the survey topic specified by the user. \textbf{(2) Personalized Reference Categorization}: our system facilitates categorization of reference papers based on user-defined criteria (\textit{e.g.}, Research Method) for personalized content organization. \textbf{(3)  Structured and Multi-modal Output}: our system generates well-organized survey papers with multiple sections and subsections. The output survey paper includes an outline diagram, as well as tables and figures from the reference papers.  \textbf{(4) Modifiable Intermediate Processes}: users can iteratively refine most steps in survey generation, including uploading local references, adjusting reference categorization results, modifying outline, and editing text content and visual elements (\textit{e.g.}, images, tables) in the generated survey paper.  \textbf{(5) Intuitive User Interface}: our system provides step-by-step guidance, clear metadata displays for references, categorization visualizations, and other user-friendly interfaces.

We comprehensively evaluate InteractiveSurvey in content quality of generated surveys, time efficiency, and usability. \textbf{Content Quality:} Evaluated by LLMs in coverage, structure, and relevance, InteractiveSurvey outperforms three mainstream LLMs in generated survey papers over 40 topics from 8 different research fields. Besides, it also outperforms state-of-the-art (SOTA) survey generation systems when generating survey papers on topics same to their released samples, without any refinement from users. \textbf{Time Efficiency:} InteractiveSurvey can produce a high-quality survey paper from scratch (with approximately 50 references) in just 35 minutes on average, using a single RTX 3090 GPU and an LLM API. \textbf{Usability:} Based on the System Usability Scale (SUS,\citet{brooke1996sus}), feedback from 34 researchers ranks our system in the highest tier with a score of 84.4/100, validating our user-friendly design. 

Our contributions can be summarized as follows: \textbf{(1): }InteractiveSurvey is an LLM-based web system that can efficiently generate high-quality survey papers for researchers. The generated survey papers outperform mainstream LLMs and SOTA survey generation systems in coverage, structure, and relevance. \textbf{(2): }As far as we know, InteractiveSurvey is the first interactive survey generation system with modifiable intermediate processes, enabling researchers to create personalized survey papers through an intuitive UI. \textbf{(3): }InteractiveSurvey is open-sourced\footnote{\href{https://github.com/TechnicolorGUO/InteractiveSurvey}{github.com/TechnicolorGUO/InteractiveSurvey}} and easy to deploy (both by direct deployment and by Docker). The backbone LLM can be replaced by any LLM via API key configuration.

\section{Related Work}

\subsection{Literature Review Generation}
Automating the summarization of multiple research papers has been of longstanding interest.
Early approaches primarily leverage multi-document summarization (MDS) techniques to generate unstructured summaries, \textit{e.g.} the related work section of a research paper \cite{hoang2010towards}. \cite{hu2014automatic} attempted to generate a related work section for a target paper given multiple reference papers as input. \cite{erera2019summarization} built the IBM Science Summarizer, which retrieves and summarizes scientific articles in computer science.

The advancements in LLMs have significantly enhanced the scope and quality of automated literature reviews. \cite{ijcai2022p591} proposed the category-based alignment and sparse transformer to generate structured summaries covering multiple research papers. ChatCite \citep{li_chatcite_2024} utilized prompt engineering to generate comparative literature summaries, \citet{susnjak_automating_2024} fine-tuned domain-specific LLMs to produce literature reviews enriched with contemporary knowledge. 

Despite these advancements, most existing studies primarily address technical challenges in literature review generation rather than producing comprehensive survey papers in practice.

\begin{figure*}
    \centering
    \includegraphics[width=1\linewidth]{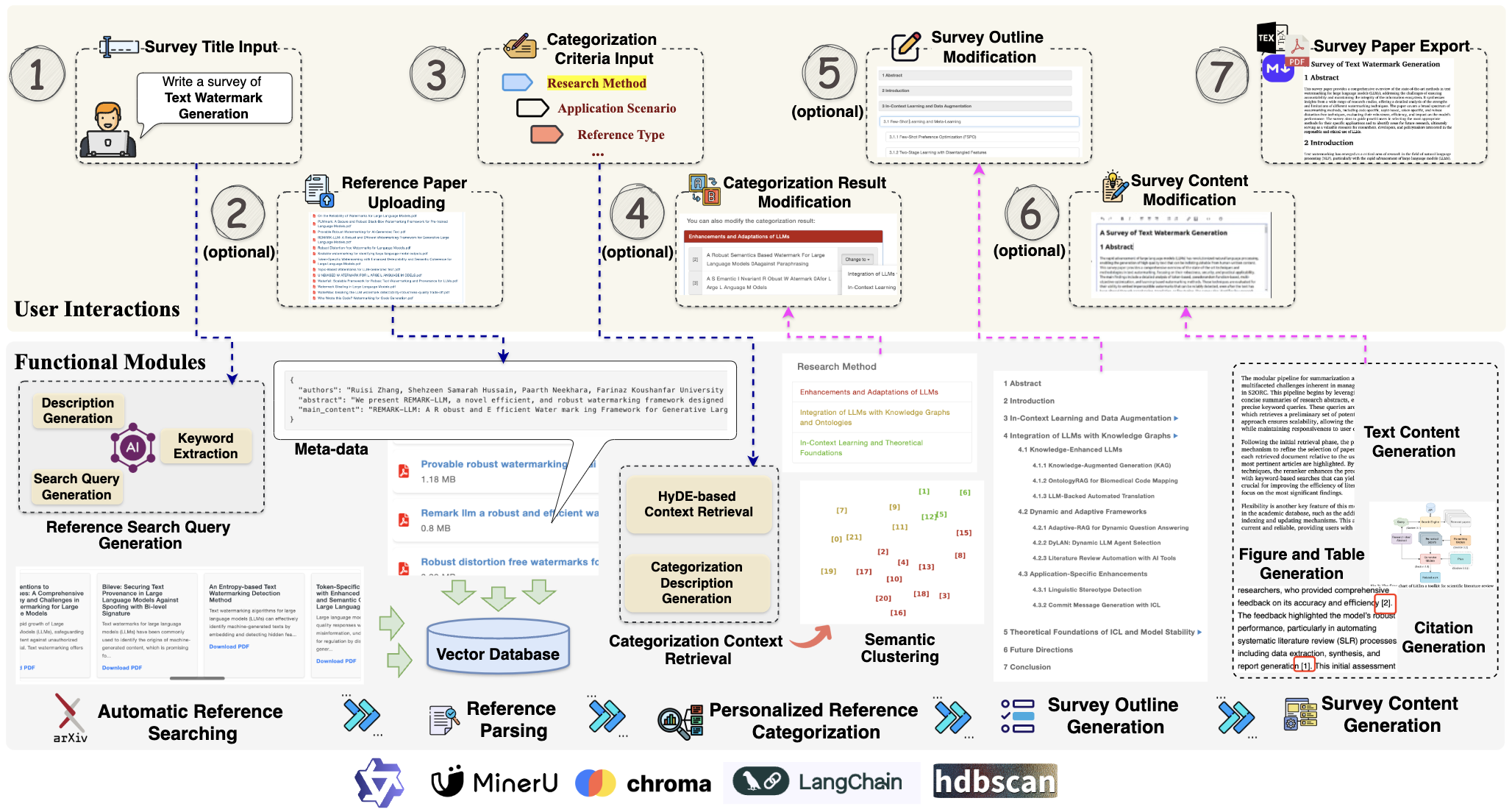}
    \caption{An Overview of InteractiveSurvey. Steps 2,4,5, and 6 in user interactions are optional.}
    \label{fig:system}
\end{figure*}

\subsection{Automated Survey Systems}

Unlike technical challenge-focused approaches, automated survey systems provide end-to-end pipelines for generating structured survey papers. Advances in RAG and document parsing techniques significantly support the implementation of these systems. LitLLM \citep{agarwal2024litllmtoolkitscientificliterature} can retrieve relevant papers and generate survey content from user-provided abstracts. HiReview \citep{hu_hireview_2024} employs hierarchical clustering on citation graphs to construct taxonomy trees for survey generation. 

Recent research has adopted more advanced strategies to produce high-quality survey papers. \citet{wang_autosurvey_2024} utilizes multi-LLM agent architectures to enable the creation of long-form content (e.g., 64k tokens). PROMPTTHEUS \citep{torres_promptheus_2024} incorporates clustering-based topic modeling to ensure structural coherence and factual accuracy. To ensure formatting consistency and adherence to academic standards,  \citet{sami_system_2024} integrated LaTeX templates in survey content generation. SurveyX \cite{liang2025surveyx} designs an end-to-end solution for automated survey generation, covering online literature search, organization, and survey content writing.

However, most systems typically limit users to title-only inputs and fixed outputs, neglecting interactive modification and refinement during generation. Consequently, users may face an all-or-nothing dilemma: either tolerate suboptimal content or restart the entire generation process.




\section{InteractiveSurvey}
\subsection{Automatic Reference Searching}

\noindent
\textbf{Online Searching} When the user inputs a topic  \( T \), our system will automatically search and download relevant references from arxiv by constructing a search query \( Q_T \) tailored for the arXiv API based on \( T \). Specifically, we first employ the LLM to generate a description $Des_{\text{T}}$ for \( T \). Then, we prompt the LLM to extract the themes $T_\text{T}$, entities $E_\text{T}$, and concepts $C_\text{T}$ from $Des_{\text{T}}$ inspired by \cite{guo_lightrag_2024}.  These components are then combined to form \( Q_T \). An example is shown in Table \ref{tab:arxiv_query_example}.

Upon obtaining \( Q_T \), we retrieve candidate references $Ref_{\text{T}}$ from arXiv. If the number of retrieved references falls below a predefined threshold MIN\_REF, we iteratively relax the search constraints by adding other related $E_\text{T}$ and $C_\text{T}$ to reformulate \( Q_T \), and repeat the search process. Finally, we truncate the results to retain at most MAX\_REF references for subsequent processing. The complete procedure is shown in Appendix \ref{Automatic Reference Searching}.

\setlength{\extrarowheight}{7pt}
\begin{table}[t] 
    \caption{The reference searching query for: \textit{A Survey of LLM in Recommendation Systems}. It returns references whose Abstract contains the following themes, entities, and concepts.}
    \centering
    \scriptsize
    \setlength{\tabcolsep}{5pt}
         \begin{tabular}{|l|}
            \hline
            \parbox{7.2cm}{ 
            \fontsize{8pt}{5pt}\selectfont \texttt{\\ \\ \textbf{Themes:} (abs:"LLM" AND abs:"recommendation") } \\
            \fontsize{8pt}{5pt}\selectfont \texttt{ \\ AND \\} \\
            \fontsize{8pt}{5pt}\selectfont \texttt{ \textbf{Entities:} (abs:"language model" } \\
            \fontsize{8pt}{5pt}\selectfont \texttt{ OR abs:"recommendation system" } \\
            \fontsize{8pt}{5pt}\selectfont \texttt{ OR abs:"contextual embedding" } \\
            \fontsize{8pt}{5pt}\selectfont \texttt{ OR abs:"semantic matching" } \\
            \fontsize{8pt}{5pt}\selectfont \texttt{ \\ AND \\} \\
            \fontsize{8pt}{5pt}\selectfont \texttt{ \textbf{Concepts:} (abs:"personalization" } \\
            \fontsize{8pt}{5pt}\selectfont \texttt{OR abs:"content understanding" }\\
            \fontsize{8pt}{5pt}\selectfont \texttt{OR abs:"collaborative filtering" }\\
            \fontsize{8pt}{5pt}\selectfont \texttt{OR abs:"matrix factorization" }\\
            }\\
            \hline
        \end{tabular}
    \label{tab:arxiv_query_example}
\end{table}

\noindent
\textbf{User Uploading}  To avoid copyright issues, we only search public arXiv papers for references. However, to accommodate users wishing to summarize local copyrighted materials, InteractiveSurvey also supports uploading such papers. 

\begin{figure}[t]
  \centering
  \includegraphics[scale=0.18]{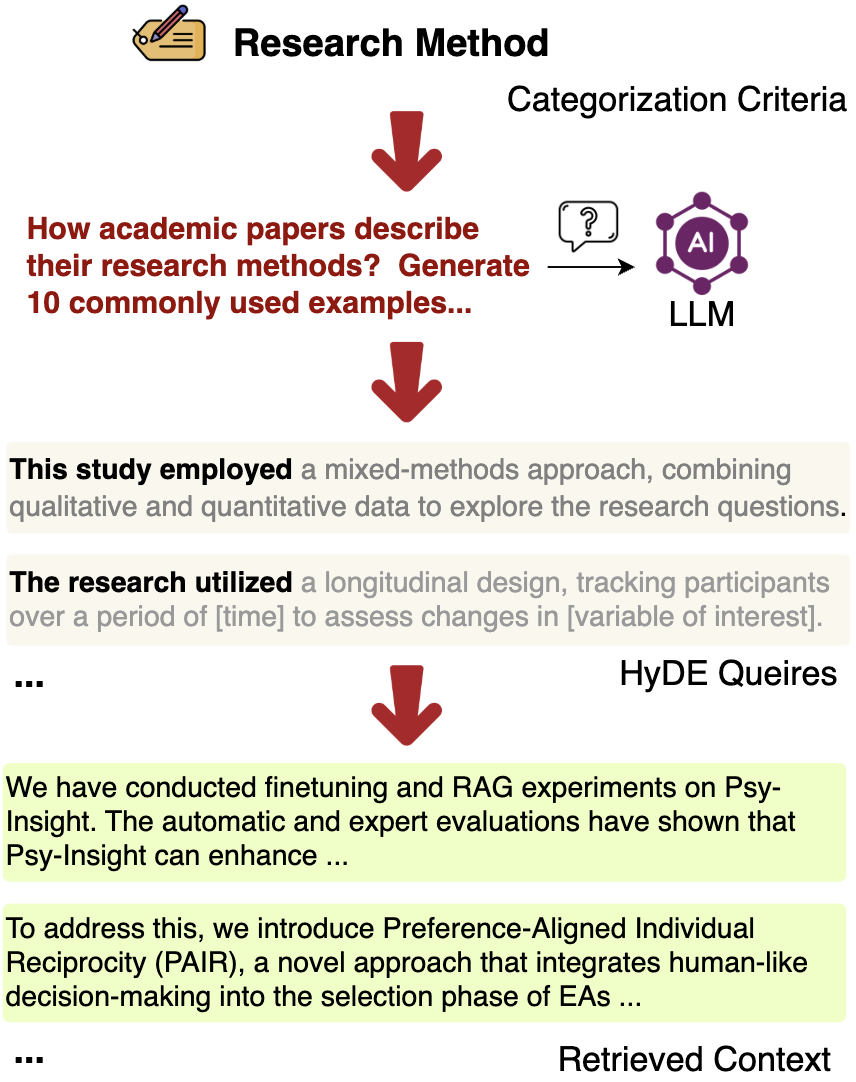}
  \caption{The HyDE Process for Retrieving Research Method Descriptions from References.}
  \label{fig:hyde}
\end{figure}

\subsection{Reference Parsing}
After collecting all reference papers from both online searching and user uploading, we parse the documents using MinerU \cite{wang2024mineruopensourcesolutionprecise}, an open-source tool that efficiently extracts and parses references (typically in PDF format) into structured Markdown files (.md). Then, our system extracts the metadata for each reference paper from the corresponding Markdown file, including title, authors, abstract, and the introduction section. The metadata is displayed on the front-end interface for user review, as shown in Figure \ref{fig:system}. Besides, the content of reference papers is stored in a vector database \(\mathcal{V}\) to facilitate subsequent processing following the procedure of RAG. Specifically, for each reference paper \(r_i \in Ref_{\text{T}} \), we split it into fixed-length chunks  \(\{d_{i}^1, d_{i}^2, \dots, d_{i}^m\}\), obtain their semantic embeddings, and store them as a collection  \(c_i\) in \(V\).

\subsection{Personalized Reference Categorization}
\label{Clustering}
Reference categorization is essential for high-quality survey papers to efficiently organize the landscape of a research area \cite{hu_hireview_2024,luo2025llm4sr}. Common categorization approaches include chronological ordering, technical taxonomy, and thematic clustering. Here, we follow the thematic clustering by letting users specify a categorization criterion $k$ and then retrieving relevant content from references for semantic clustering.

\subsubsection{Categorization Context Retrieval}
\label{Categorization Context Retrieval}

\noindent
\textbf{HyDE-based Context Retrieval} To facilitate accurate retrieval, we employ Hypothetical Document Embeddings (HyDE, \citet{gao2023precise}), which generates potential pseudo-descriptions to the categorization criterion $k$ as queries for semantic matching. Specifically, we prompt the LLM to generate 10 different HyDE queries for $k$. For each collection \(c_i\) corresponding to $r_i$, all generated queries are used in parallel to retrieve the most relevant chunks through semantic matching. Then, we merge and de-duplicate all the retrieved chunks, yielding a consolidated set of retrieved context $C_i^k$. An example is shown in Figure \ref{fig:hyde}.

\begin{figure}[t]
  \centering
  \includegraphics[scale=0.14]{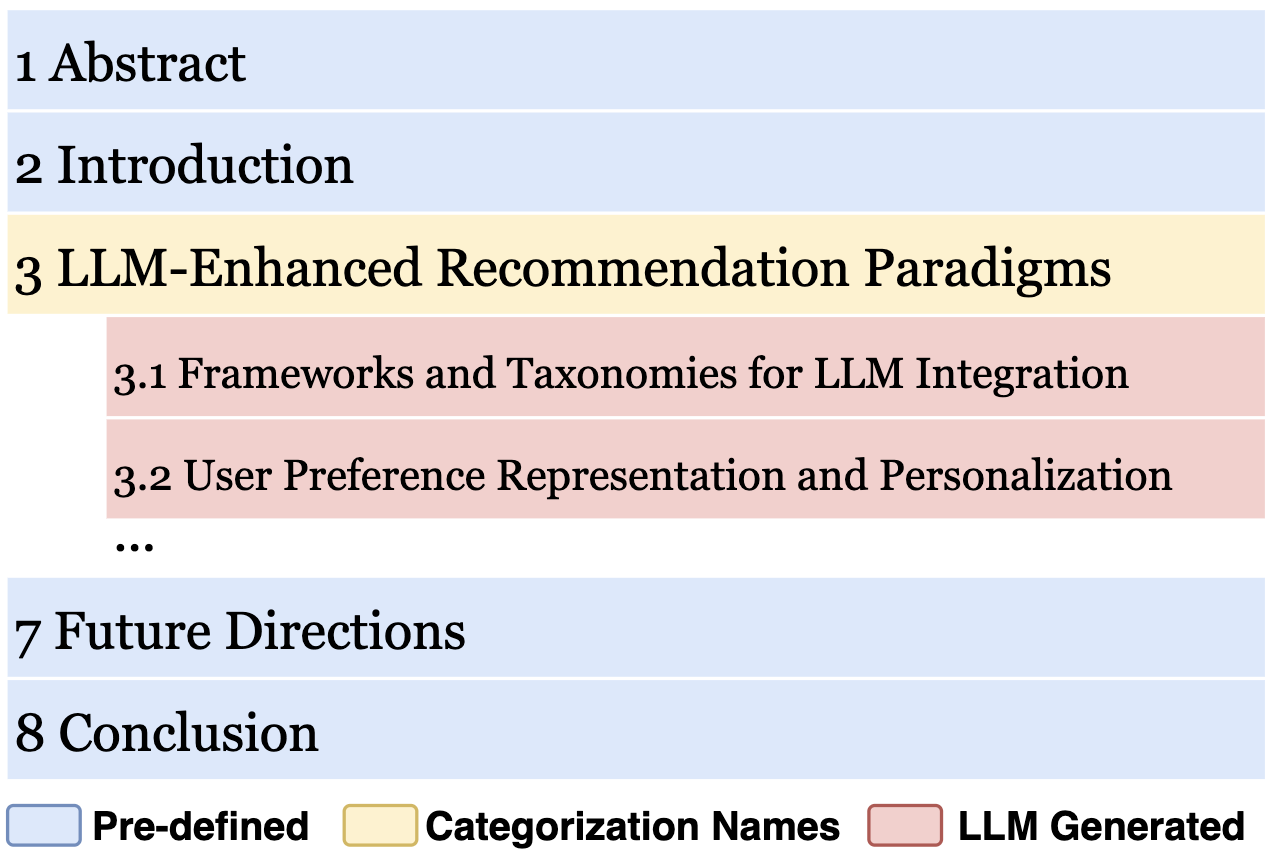}
  \caption{The pre-defined section names and the categorization names are as section titles, sub-section titles are generated by the LLM.}
  \label{fig:outline}
\end{figure}

\noindent
\textbf{Categorization Description Generation} The retrieved context $C_i^k$ is then fed to the LLM to generate the description $d_i^k$ for \(r_i\), best representing \(r_i\)'s relevant content to the categorization criterion $k$.

\noindent
\textbf{Semantic Clustering}
After generating descriptions \( \{d_1^k, d_2^k, \dots, d_n^k\} \) for all reference papers, we perform an agglomerative semantic clustering on these descriptions and present the results to users, including (1) a t-SNE visualization for cluster spatial relationships, (2) categorization names $\mathcal{N}$ describing references in clusters, and (3) complete sets of categorized reference papers. The detailed clustering procedure is shown in Appendix \ref{appendix_clustering}. Besides, InteractiveSurvey supports interactive refinements: users may manually adjust the clustering results by moving references among clusters to better meet their analytical needs, as shown in Steps 3 and 4 in Figure \ref{fig:system}.

\subsection{Survey Outline Generation}

To facilitate the generation of high-quality survey content, we construct a three-level hierarchical outline $O$ (Figure \ref{fig:outline}) including common section titles (\textit{e.g.}, Abstract, Introduction, Conclusion) in most survey papers, the categorization names $\mathcal{N}$ as section titles, and the LLM-generated sub-section titles from reference descriptions obtained in Section \ref{Clustering}. The generated outline also allows for iterative refinement through manual edits, as shown in Step 5 in Figure \ref{fig:system}.

To ensure the LLM-generated outline follows the hierarchical structure format:
\[O = \{(l_1, t_1), (l_2, t_2), \dots, (l_m, t_m)\} \]
where $l_i$ is the hierarchical level, and $t_i$ is the section/sub-section title, we first provide the LLM with the outline template as well as partial content of $t_i$ in the prompt, and then ask it to fill in the blanks. We found that this method achieves significantly better format following than direct end-to-end generation, which can be useful when deploying InteractiveSurvey with less powerful LLMs.

\subsection{Survey Content Generation}
\label{subsec:section-generation}

Based on the outline above, InteractiveSurvey generates structured and multi-modal survey content, including text content, images/figures, and citations. The generated content is fully editable and exportable in multiple formats, including PDF, Markdown, and LaTeX, as shown in Figure \ref{fig:system}.

\noindent
\textbf{Text Content Generation}
We adopt a bottom-up approach to generate text content section by section. For sections titled with categorization names, we first use their respective sub-section titles as queries to retrieve reference content from the vector database $\mathcal{V}$. This retrieved content serves as the prompt to the LLM to generate the sub-section content. Subsequently, we employ the LLM to produce a summary of the sub-section content. The summary and the sub-section content together form the section content. For pre-defined title sections such as \textit{Abstract}, \textit{Introduction}, \textit{Future Directions}, and \textit{Conclusion}, we use the already-generated section content as the input and prompt the LLM to generate their section content with different instructions. The detailed process is shown in Appendix \ref{Text Content Generation}.

\noindent
\textbf{Figure and Table Generation}
To enable multi-modal output, we incorporate two types of figures and tables into the generated content: (1) the image of the structure visualization\footnote{By Graphviz at \href{https://graphviz.org/}{https://graphviz.org/}} of the survey paper outline, and (2) figures/tables retrieved from reference papers. We conduct semantic matching between sentences in the generated survey paper and the captions of figures/tables in the references. Those with matching scores above the pre-defined threshold are incorporated into the generated survey paper, along with citation information.

\noindent
\textbf{Citation Generation}
To facilitate researchers' reading experience, we also generate citations in the survey paper linking back to the reference papers. Inspired by the academic writing of human and existing studies \cite{gao2023enabling,wang_autosurvey_2024}, we retrieve relevant chunks from all reference papers for each sentence in the generated survey and apply an adaptive semantic similarity threshold to incorporate an appropriate number of citations. The details are shown in Appendix \ref{Citation Generation}.

\section{Evaluation}

\setlength{\extrarowheight}{1pt}
\begin{table*}[t]
\caption{Comparison on survey content quality among different LLMs and InteractiveSurvey (with Qwen2.5-72B API). Prompt means we directly prompt the LLMs to generate a survey paper on the given topic. Abstract means we input the abstracts of the search reference papers aligned with the given topic.}
\centering
\scriptsize
\setlength{\tabcolsep}{5pt}
\begin{tabular}{|c|c|cc|cc|cc|c|}
\hline
\multirow{2}{*}{\textbf{Aspects}} & \multirow{2}{*}{\textbf{LLM Judges}} & \multicolumn{2}{c|}{\textbf{Qwen2.5-72B}} & \multicolumn{2}{c|}{\textbf{GPT-4o}} & \multicolumn{2}{c|}{\textbf{DeepSeek-R1}} & \multirow{2}{*}{\parbox{3cm}{\centering \textbf{InteractiveSurvey}\\ \textbf{(Qwen2.5-72B)} }} \\
\cline{3-8}
 & & \textbf{Prompt} & \textbf{Abstract} & \textbf{Prompt} & \textbf{Abstract} & \textbf{Prompt} & \textbf{Abstract} & \\
\hline
\multirow{4}{*}{Coverage $\uparrow$} 
 & Qwen2.5-72B & 4.12 & 4.10 & 4.46 & 4.18 & 4.22 & 4.37 & \textbf{4.58} \\
 & GPT-4o & 4.05 & 4.03 & 4.34 & 4.63 & 4.08 & 4.13 & \textbf{4.67}\\
 & DeepSeek-R1 & 4.10 & 4.38 & 4.40 & 4.39 & 4.30 & \textbf{4.83} & 4.43 \\
 \cline{2-9}
 & \cellcolor{gray!20}Avg. & \cellcolor{gray!20}4.09 & \cellcolor{gray!20}4.17 & \cellcolor{gray!20}4.40 & \cellcolor{gray!20}4.40 & \cellcolor{gray!20}4.20 & \cellcolor{gray!20}4.44 & \cellcolor{gray!20}\textbf{4.56} \\
\hline
\multirow{4}{*}{Structure $\uparrow$} 
 & Qwen2.5-72B & 4.25 & 4.23 & 4.48 & 4.21 & 4.32 & 4.24 & \textbf{4.50} \\
 & GPT-4o & 4.35 & 4.21 & 4.56 & 4.44 & 4.30 & 4.25 & \textbf{4.73} \\
 & DeepSeek-R1 & 4.03 & 4.08 & 4.46 & 4.55 & 4.43 & 4.45 & \textbf{4.55} \\
 \cline{2-9}
 & \cellcolor{gray!20}Avg. & \cellcolor{gray!20}4.21 & \cellcolor{gray!20}4.17 & \cellcolor{gray!20}4.50 & \cellcolor{gray!20}4.40 & \cellcolor{gray!20}4.35 & \cellcolor{gray!20}4.31 & \cellcolor{gray!20}\textbf{4.59} \\
\hline
\multirow{4}{*}{Relevance $\uparrow$} 
 & Qwen2.5-72B & 4.78 & 4.83 & 4.78 & 4.60 & 4.78 & 4.60 & \textbf{4.85}\\
 & GPT-4o & 4.55 & 4.65 & 4.82 & 4.90 & 4.70 & 4.33 & \textbf{4.95}\\
 & DeepSeek-R1 & 4.83 & 4.75 & 4.78 & 4.75 & 4.77 & 4.75 & \textbf{4.83} \\
\cline{2-9}
 & \cellcolor{gray!20}Avg.  & \cellcolor{gray!20}4.72 & \cellcolor{gray!20}4.74 & \cellcolor{gray!20}4.80 & \cellcolor{gray!20}4.75 & \cellcolor{gray!20}4.75 & \cellcolor{gray!20}4.56 &\cellcolor{gray!20}\textbf{4.88} \\
\hline
\end{tabular}
\label{tab:comparison with LLMs}
\end{table*}

\subsection{Content Quality}

\noindent
\textbf{Evaluation Metrics} Following existing survey generation systems \cite{wang_autosurvey_2024,liang2025surveyx}, we employ LLMs as judges \cite{li2024llms} to assess the survey papers in \textit{Coverage}: the extent to which the survey encapsulates all aspects of the topic; \textit{Structure}: the logical organization and coherence of each section; and \textit{Relevance}: how well the content aligns with the user-input topic. The LLM prompt for evaluation is provided in Appendix \ref{evaluation prompt}.

\noindent
\textbf{Comparison with Different LLMs} We compare InteractiveSurvey with various types of LLMs (\textit{i.e.}, GPT-4o \cite{hurst2024gpt}, DeepSeek-R1 \cite{guo2025deepseek}, and Qwen2.5-72b \cite{yang2024qwen2} on their survey generation abilities. The comparison setting is in Appendix \ref{llm setting}. As shown in Table \ref{tab:comparison with LLMs}, even if different LLMs were used as judges across the three evaluation aspects, our approach consistently outperformed mainstream LLMs in most cases, underscoring the high quality of our generated survey content. We also observed marginal differences in quality scores between using only the title versus the full abstract as input for LLM-generated survey papers. This suggests that most LLMs are constrained by context window limitations to effectively retain and utilize information, even when more extensive context is provided \cite{chen2023extending}.

\noindent
\textbf{Comparison with SOTA Methods} In addition to LLMs, we also compare InteractiveSurvey with two SOTA survey generation methods: AutoSurvey \cite{wang_autosurvey_2024} and SurveyX \cite{liang2025surveyx} on the generated survey paper samples. The comparison setting is in Appendix \label{sota setting}. As shown in Table \ref{tab:survey_comparison}, our method achieves the highest scores across all evaluation metrics on average, which also demonstrates that InteractiveSurvey outperform SOTA methods in survey content quality.

\setlength{\extrarowheight}{1pt}
\begin{table}[t]
    \caption{Comparison between InteractiveSurvey and SOTA methods on survey samples.}
    \centering
    \scriptsize
    \setlength{\tabcolsep}{2pt}
    \begin{tabular}{|c|c|ccc|}
        \hline
        \multirow{2}{*}{\textbf{Aspects}} & \multirow{2}{*}{\textbf{LLM Judges}} & \multicolumn{3}{c|}{\textbf{Survey Generation Methods}} \\
        \cline{3-5}
        & & \textbf{AutoSurvey} & \textbf{SurveyX} & \textbf{InteractiveSurvey} \\
        \hline
        \multirow{4}{*}{Coverage $\uparrow$} 
        & Qwen2.5-72B       & 4.67 & 4.37 & \textbf{4.82} \\
        & GPT-4o            & 4.67 & 4.09 & \textbf{4.68} \\
        & DeepSeek-R1       & 4.00 & 4.18 & \textbf{4.29} \\
        \cline{2-5}
        & \cellcolor{gray!20}Avg.              & \cellcolor{gray!20}4.44 & \cellcolor{gray!20}4.21 & \cellcolor{gray!20}\textbf{4.61} \\
        \hline
        \multirow{4}{*}{Structure$\uparrow$ } 
        & Qwen2.5-72B       & 4.67 & 4.71 & \textbf{4.76} \\
        & GPT-4o            & 4.33 & 4.26 & \textbf{4.40} \\
        & DeepSeek-R1       & \textbf{4.67} & 3.97 & 4.64 \\
        \cline{2-5}
        & \cellcolor{gray!20}Avg.              & \cellcolor{gray!20}4.56 & \cellcolor{gray!20}4.31 & \cellcolor{gray!20}\textbf{4.60} \\
        \hline
        \multirow{4}{*}{Relevance $\uparrow$} 
        & Qwen2.5-72B       & 4.67 & 4.89 & \textbf{4.94} \\
        & GPT-4o            & 4.67 & 4.26 & \textbf{4.69} \\
        & DeepSeek-R1       & 4.67 & 4.29 & \textbf{4.78} \\
        \cline{2-5}
        & \cellcolor{gray!20}Avg.              & \cellcolor{gray!20}4.67 &\cellcolor{gray!20} 4.48 & \cellcolor{gray!20}\textbf{4.80} \\
        \hline
    \end{tabular}
    \label{tab:survey_comparison}
\end{table}

\subsection{Time Efficiency}
\begin{figure}[t]
    \centering
  \includegraphics[scale=0.16]{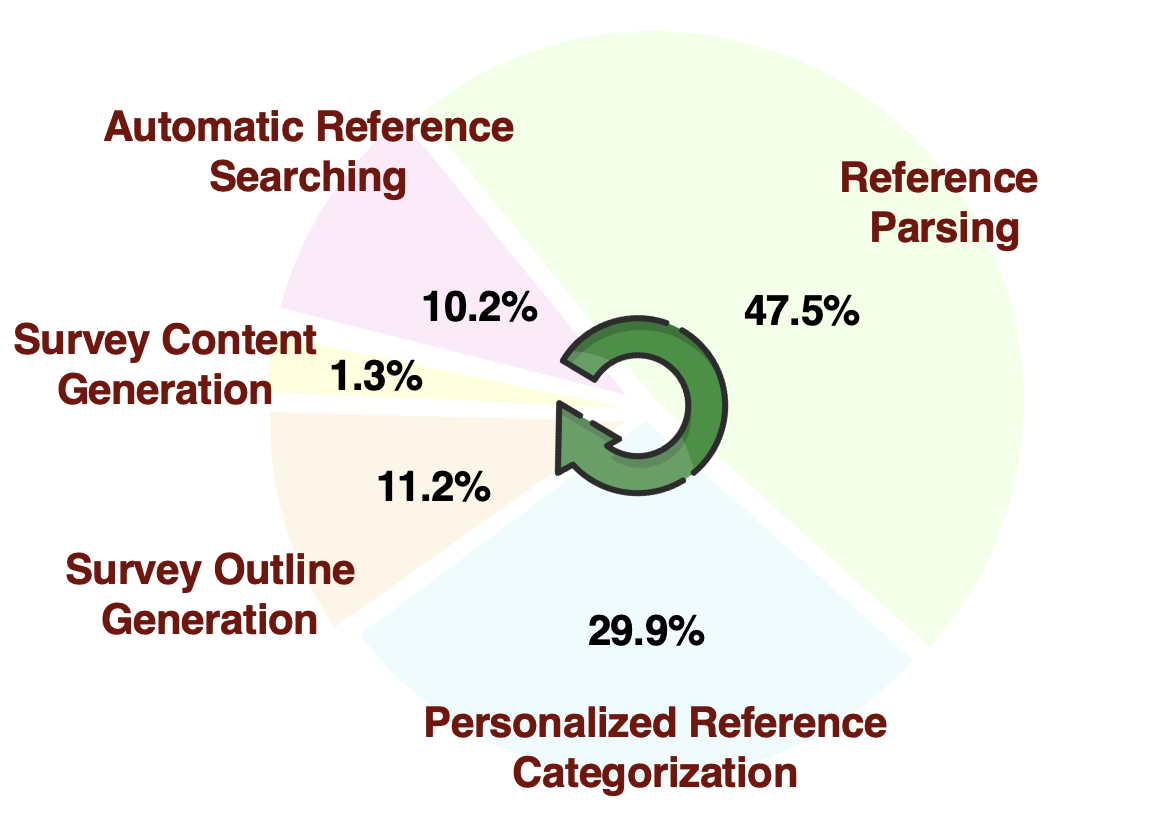}
  \caption{Time cost distribution in InteractiveSurvey}
  \label{fig:timepie}
\end{figure}

In addition to content quality, we evaluated the time efficiency of InteractiveSurvey. We use the 40-topics reference collection described in Appendix \ref{llm setting} (with an average of 45.2 references per topic) to generate 40 survey papers with default settings for user interactions. The experiments were conducted on the following hardware/software configuration: (1) GPU: NVIDIA GeForce RTX 3090 (for reference parsing and categorization), (2) CPU: AMD Ryzen 9 5900X 12-Core Processor (for web system deployment), (3) LLM API: qwen2.5-72b-instruct (for RAG support, outline generation, and survey content creation). The average time required to generate one survey paper is 2,077.8 seconds (~35 minutes). Figure \ref{fig:timepie} shows the detailed time distribution across different processing stages.

Our analysis shows that the most time-intensive steps are Reference Parsing and Personalized Reference Categorization, which may be constrained by our GPU's computational power. Although Personalized Reference Categorization also includes the time for LLM-based RAG, we anticipate that with more advanced hardware, the total processing time could be reduced to under 30 minutes.

\subsection{User Study with SUS Test}
We also conducted a user study involving 34 participants, including PhD students, research assistants, and postdoctoral fellows. They are asked to experience and assess InteractiveSurvey by the System Usability Scale (SUS), a widely adopted questionnaire for measuring system usability. Following \cite{sauro2012quantifying}, we converted the raw scores into a normalized 100-point scale. InteractiveSurvey achieved an outstanding score of 84.4, placing it in the A+ tier (the highest grade) for usability. This result demonstrates that InteractiveSurvey is really easy to use for researchers. The detailed scale and the calculation method are provided in Appendix \ref{sus}.

\section{Conclusion}
In this paper, we introduce InteractiveSurvey, a web-based system powered by large language models (LLMs) that efficiently generates high-quality survey papers. To the best of our knowledge, it is the first interactive platform that enables researchers to refine intermediate outputs via an intuitive UI, allowing for personalized survey creation. Future work will involve collecting user feedback post-deployment to enhance functionality, UI design, and evaluation. Additionally, we plan to explore multilingual survey generation to broaden the system’s applicability.

\bibliography{main}
\bibliographystyle{acl_natbib}

\appendix
\clearpage
\section{Pseudo Codes}
\subsection{Pseudo Code of Automatic Reference Searching}
\label{Automatic Reference Searching}
\begin{algorithm}[h]
\caption{Automatic Reference Searching} 
\label{alg1}
\small
\begin{algorithmic}[1]
\REQUIRE $T$, MAX\_REF, MIN\_REF, MAX\_TRY
\ENSURE $Ref_{\text{T}} = \{r_1,r_2, ..., r_n\}$ 
\STATE $Des_{\text{T}} \gets \text{LLM}(T)$  
\STATE $T_{\text{T}}, E_{\text{T}}, C_{\text{T}}\gets \text{LLM}_{\text{extract}}(T, Des_{\text{T}})$
\STATE $Q_{\text{T}} = T_{\text{T}} \land E_{\text{T}} \land C_{\text{T}}$
\WHILE{{search times $\leq$ MAX\_TRY}}
    \STATE $Result \gets$  Search($Q_{\text{T}}$)
    \STATE $Ref_{\text{T}} \gets Ref_{\text{T}} \cup Result$
    \IF{Size($Ref_{\text{T}}$) < MIN\_REF }
        \STATE $Q_T \gets $ Loose($Q_{\text{T}}|T_{\text{T}}$)
    \ENDIF
\ENDWHILE
\STATE $Ref_{\text{T}} \gets Ref_{\text{T}}[:\text{MAX\_REF}]$
\end{algorithmic}   
\end{algorithm}

\subsection{Pseudo Code of Text Content Generation}
\label{Text Content Generation}
\begin{algorithm}[h]
\small
\caption{Text Content Generation}
\label{alg2}
\begin{algorithmic}[1]
\REQUIRE $O = \{(l_1, t_1), \dots, (l_m, t_m)\}, \mathcal{V}$
\ENSURE $S = \{(l_1, t_i, s_1), \dots, (l_m, t_i, s_m)\}$ 
\FOR{Cluster Name $(l_i, t_i) \in O$ }
    \STATE \texttt{//in parallel}
    \FOR{each sub-section title $t_k$ under $t_i$}
        \STATE  $ sub\_context_j \gets$ Retrieve($t_k,\mathcal{V}$)
        \STATE  $ s_k \gets$ RAG($t_k, sub\_context_j$)
    \ENDFOR
    \STATE $sum_i$ $\gets$ $\text{LLM}_{\text{Sum}}$(subsections $s_1, ..., s_n$)
    \STATE $s_i$ $\gets$ $\text{LLM}_{\text{Merge}}$($t_i, sum_i, s_1, ..., s_n$)
\ENDFOR
\FOR{Pre-defined section titles $(l_j, t_j) \in O$ }
   \STATE \texttt{//in parallel}
   \STATE $s_i$ $\gets$ LLM(generated $s_1, ..., s_k$)
\ENDFOR
\STATE $S \gets \{(l_1, t_i, s_1), \dots, (l_m, t_i, s_m)\}$ 
\end{algorithmic}   
\end{algorithm}

\section{Details of Semantic Clustering}
\label{appendix_clustering}

The modularized clustering process comprises the following sequential steps:

\begin{enumerate}

    \item \textbf{Semantic Embedding}: \( \{d_1^k, d_2^k, \dots, d_n^k\} \) corresponding to each reference \(r_i\) and criteria \(k\) are transformed into $m$-dimensional semantic representations \( \mathbf{V} = \{v_1, v_2, \dots, v_n\} \) using an embedding model \cite{nussbaum2024nomic} \( \phi: D \rightarrow \mathbb{R}^m \). 

    \item \textbf{Dimensionality Reduction}: To facilitate efficient clustering and visualization, \( \mathbf{V} = \{v_1, v_2, \dots, v_n\} \) are projected into a lower-dimensional space using the Uniform Manifold Approximation and Projection (UMAP, \citet{mcinnes2018umap}).
    \[
    \mathbf{U} = \text{UMAP}(\mathbf{V}) \rightarrow \mathbb{R}^q, \quad q \ll m,
    \]
   where \( \mathbf{U} = \{u_1, u_2, \dots, u_n\} \) are the reduced-dimensional representations, $q$ is a hyper-parameter indicating the target dimension.

    \item \textbf{Adaptive Clustering}: We perform hierarchical clustering on the dimensionally reduced representations \( \mathbf{U} \) using the HDBSCAN algorithm \cite{mcinnes2017hdbscan}. Given the inherent variability in survey topics, reference papers, and input categorization criteria $k$, there is no universally optimal number of clusters. To address this, we first specify several candidate cluster numbers (\textit{e.g.}, 3–6) and then the Silhouette score \cite{rousseeuw1987silhouettes} is used to select the number with the highest score, indicating the most coherent clustering structure.

    \item \textbf{Categorization Name Generation}: After we obtain the clustering results, for each cluster \( C_j \), we aggregate the descriptions (generated in Section \ref{Categorization Context Retrieval}) for all reference papers within \( \{d_i \mid r_i \in C_j\} \) and provide them to the LLM to generate a representative categorization name \( N_j \) to captures the cluster's thematic focus relative to criteria $k$. Notably, the LLM here is configured to generate all cluster names for all $L$ clusters simultaneously to ensure their coherence in expression.
    
    \[
    \mathcal{N} = \text{LLM}\left( \bigcup_{j=1}^{L} \{ d_i \mid r_i \in C_j \} \right )
    \]
    
    where \( \mathcal{N} = \{N_1, N_2, \dots, N_L\} \) represents the set of generated cluster names.
    
\end{enumerate}

\section{Citation Generation}
\label{Citation Generation}

For each sentence \(s_i \in \{s_1, s_2, \dots, s_n\}\) in the generated survey paper, we calculate the cosine similarity $\mathrm{sim}_{i,j}$ for semantic representations of each sentence $s_i$ to each chunk \(c_j\) in the reference vector database $\mathcal{V}$. A sentence-chunk pair \(\bigl(s_i, c_j\bigr)\) is assigned a citation if \(\mathrm{sim}_{i,j} \geq \tau\), where \(\tau\) is the semantic similarity threshold. As a constant threshold may result in localized overcitation or undercitation within specific passages, we adopt an adaptive approach that adjusts  \(\tau\) based on the global distribution of similarity scores:
\[
\tau = \max\!\bigl(\tau_{\mathrm{static}},\;\mu + k\,\sigma\bigr),\]
where $\tau_{\mathrm{static}}$ is the initialized static threshold, \(\mu\) and \(\sigma\) denote the mean and standard deviation of similarity scores across all $\mathrm{sim}_{i,j}$ pairs, and \(k\) is a hyper-parameter to adjust the strictness of the threshold.

\section{Details of the SUS Test}
\label{sus}
In our user study, 34 participants were asked to fill in the questionnaire with items in Table \ref{sus table} to score each item with 1 (Strongly Disagree) to 5 (Strongly Agree). Then, we calculate the average scores for each item, and convert the raw scores into a normalized 100-point scale by:
$$
\text{Score} = 4*
\begin{aligned}[t]
[ & (s_1+s_3+s_5+s_7+s_9 - 5) + \\
  & (25-s_2-s_4-s_6-s_8-s_{10}) ]
\end{aligned}
$$
where, $s_i$ is the average score of $i$-\textit{th} item. Our score 84.4 falls into the A+ tier with the range (84.1-100) according to \cite{sauro2012quantifying}. The raw scores are released on our github repository.

\setlength{\extrarowheight}{2pt}
\begin{table}[h]
\centering
 \setlength{\tabcolsep}{2pt}
 \renewcommand{\arraystretch}{1.2}
\scriptsize
\caption{Items in the SUS \cite{brooke1996sus}}
\begin{tabular}{|cl|}
\hline
1 &\parbox{7cm}{I think that I would like to use this system frequently.}\\
2 &\parbox{7cm}{I found the system unnecessarily complex.}\\
3 &\parbox{7cm}{I thought the system was easy to use.}\\
4 &\parbox{7cm}{I think that I would need the support of a technical person to be able to use this system.}\\
6 &\parbox{7cm}{I found the various functions in this system were well integrated.}\\
6 &\parbox{7cm}{I thought there was too much inconsistency in this system.}\\
7 &\parbox{7cm}{I would imagine that most people would learn to use this system very quickly.}\\
8 &\parbox{7cm}{I found the system very cumbersome to use.}\\
9 &\parbox{7cm}{I felt very confident using the system.}\\
10 &\parbox{7cm}{I needed to learn a lot of things before I could get going with this system.}\\
\hline
\end{tabular}
\label{sus table}
\end{table}

\section{Experiment Settings for Content Quality Evaluation}
\label{evaluation}

\subsection{LLM Prompts for Evaluation}
\label{evaluation prompt}
We evaluate each generated survey paper by employing LLMs to score it using the prompt in Table \ref{prompt}. \texttt{[TOPIC]} represents the survey title provided by the user, while \texttt{[SURVEY CONTENT]} corresponds to the textual content of the generated survey paper. The remaining placeholders are filled in with the content quality criteria from \cite{wang_autosurvey_2024}. Notably, some of the generated survey papers contain images, and all the LLMs used for evaluation support the uploading of PDF files. Therefore, we directly uploaded the PDFs of the generated survey papers instead of inserting the \texttt{[SURVEY CONTENT]} text during the actual evaluation process. 

\begin{table}[t] 
    \centering
    \scriptsize
    \setlength{\tabcolsep}{5pt}
    \caption{LLM prompts for evaluation}
    \label{prompt}
    \begin{tabular}{|l|}
        \hline
         \parbox{7.2cm}{ 
         \fontsize{8pt}{5pt}\selectfont \texttt{\\ \\ Here is an academic survey about the topic "[TOPIC]":} \\
         \fontsize{8pt}{5pt}\selectfont \texttt{\\ --- \\ } \\
         \fontsize{8pt}{5pt}\selectfont \texttt{[SURVEY CONTENT]} \\
         \fontsize{8pt}{5pt}\selectfont \texttt{\\ --- \\ } \\
         \fontsize{8pt}{5pt}\selectfont \texttt{<instruction>} \\
         \fontsize{8pt}{5pt}\selectfont \texttt{ Please evaluate this survey about the topic "[TOPIC]" based on the criteria above provided below, and give a score from 1 to 5 according to the score description: \\ } \\
         \fontsize{8pt}{5pt}\selectfont \texttt{\\ --- \\ } \\
         \fontsize{8pt}{5pt}\selectfont \texttt{ Criterion Description: [Criterion Description]} \\
         \fontsize{8pt}{5pt}\selectfont \texttt{\\ --- \\ } \\
         \fontsize{8pt}{5pt}\selectfont \texttt{ Score 1 Description: [Score 1 Description] } \\
         \fontsize{8pt}{5pt}\selectfont \texttt{ Score 2 Description: [Score 2 Description] } \\
         \fontsize{8pt}{5pt}\selectfont \texttt{ Score 3 Description: [Score 3 Description] } \\
         \fontsize{8pt}{5pt}\selectfont \texttt{ Score 4 Description: [Score 4 Description] } \\
         \fontsize{8pt}{5pt}\selectfont \texttt{ Score 5 Description: [Score 5 Description] } \\
         \fontsize{8pt}{5pt}\selectfont \texttt{\\ --- \\ } \\
         \fontsize{8pt}{5pt}\selectfont \texttt{Return the score without any other information: }\\
         } \\
        \hline
    \end{tabular}
\end{table}

\subsection{Settings of Comparison with Different LLMs}
\label{llm setting}
For comprehensive evaluation, we select 40 topics from 8 different research fields\footnote{Computer Science, Mathematics, Physics, Statistics, Electrical Engineering and Systems Science, Quantitative Biology, Quantitative Finance, and Economics} on arXiv and automatically search references through our system for survey paper generation. The topics, reference papers, and generated survey papers are also released on our github repository.
Considering the input context windows limits of LLMs, we employ two approaches for them to generate survey papers:  (1) \textit{Prompt}: We directly prompt the LLMs to generate a survey paper on a given topic without any additional input; and  (2) \textit{Abstract}: We input all the abstracts of the search reference papers aligned with the given topic as the prompt. 

\subsection{Settings of Comparison with SOTA Methods}
\label{sota setting}
As the source code of AutoSurvey is continuously
iterated and SurveyX hasn’t released its implementation, we compare survey papers generated by us with the survey samples they released\footnote{\href{http://github.com/AutoSurveys/AutoSurvey/tree/main/examples}{AutoSurvey examples} and \href{http://github.com/IAAR-Shanghai/SurveyX/tree/main/examples}{SurveyX examples}}.
Specifically, we input the identical survey titles to their released survey papers into InteractiveSurvey and automatically search reference papers and generate survey papers. These papers and the survey samples they released were jointly evaluated and compared by LLM judges.
Samples from AutoSurvey and SurveyX are generated by GPT-4o according to their description, survey generated by InteractiveSurvey is with Qwen2.5-72b API.

\end{document}